# Increase of SERS Signal Upon Heating or Exposure to a High-Intensity Laser Field: Benzenethiol on an AgFON Substrate

*Roshan L. Aggarwal[a*], Lewis W. Farrar[a], and Semion K. Saikin[b]*

[a]MIT Lincoln Laboratory, Lexington, MA 02420-9108, USA

[b]Department of Chemistry and Chemical Biology, Harvard University, Cambridge, MA 02138, USA

ABSTRACT The surface-enhanced Raman scattering (SERS) signal from an AgFON plasmonic substrate, recoated with benzenethiol, was observed to increase by about 100% upon heating for 3.5 min at 100º C and 1.5 min at 125º C. The signal intensity was found to increase further by about 80% upon a 10-s exposure to a high-intensity (3.2 kW/cm$^2$) 785-nm cw laser, corresponding to 40 mW in a 40±5-µm diameter spot. The observed increase in the SERS signal may be understood by considering the presence of benzenethiol molecules in an intermediate or 'precursor' state in addition to conventionally ordered molecules forming a self-assembled monolayer. The increase in the SERS signal arises from the conversion of the molecules in the precursor state to the chemisorbed state due to thermal and photo-thermal effects.

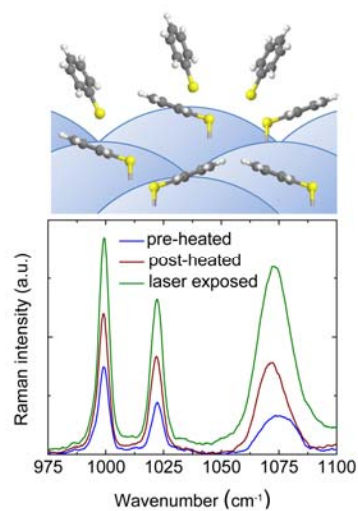

## INTRODUCTION

The phenomenon of surface-enhanced Raman scattering (SERS)[1-4] possesses a great potential in sensing and recognition[5-7] of molecular structures up to a single-molecule level.[8,9] In the SERS setup plasmon excitations supported by the noble metal substrate focus the probing laser field below the diffraction limit and also enhance the spontaneous emission from the analyte molecules adsorbed on the surface. This results in the amplification of the Raman signal on average by more than a million times as compared to the conventional setup of Raman spectroscopy.[10,11] In addition, the response from the molecules is modified -- vibrational peaks are shifted and relative intensities are changed. The latter effects is associated with the adsorption of the molecules on the metal surface.[12-15]

Benzenethiol (PhSH) adsorbed on a plasmonic substrate is a canonical system to study SERS. As many other thiolates these molecules form ordered monolayers on noble metal surfaces,[16] which allows for quantitative estimations of molecular surface density, and prediction of molecular orientation on the surface -- crucial parameters for the calculation of Raman signal enhancement. Benzenethiol can be chemisorbed on a substrate from a gas phase or a solution breaking SH bond and releasing hydrogen. The conventional procedure for the PhSH coating of noble surfaces starts with a clean plasmonic surface, which is natural for lab experiments, but can be impractical for possible sensor applications. In this study we intentionally break the conventional paradigm and explore the spectra when the plasmonic substrate is recycled. By monitoring Raman spectra of PhSH molecules on the surface we address the fundamental issue of molecular adsorption and reorganization on the surface.

Recently, an increase (> 500%) in the SERS intensity from a self-assembled monolayer (SAM) of benzenethiol on a Klarite SERS substrate was reported by Hugall *et al.* using a $CO_2$ snow jet that induces molecular redistribution and surface morphology on the substrate.[17] In this paper, we report an increase (~ 100%) in the SERS intensity from recoated benzenethiol on a silver film over nanospheres (AgFON)[10,18] substrate using heating (3.5 min at 100º C, and 1.5 min at 125º C) and/or 10-s laser exposure to a high-intensity (3.2 kW/cm$^2$) 785-nm cw pump laser, corresponding to 40 mW pump power in a 40±5 μm diameter spot. The observed increase in the intensity may be caused by the conversion of the benzenethiol molecules in the precursor state to adsorbed state due to thermal effect for the case of heating and photo-thermal effect for the case of laser exposure.

EXPERIMENTAL

The AgFON SERS substrate was fabricated at Northwestern University using $SiO_2$ nanospheres and 200-nm thick Ag film following the procedure described previously.[19] The SERS substrate was initially coated with benzenethiol in November 2009 according to the conventional procedure: the AgFON substrate was placed in 4-mM ethanol solution of benzenethiol for 2 hours and then gently rinsed in neat ethanol for 1 min, followed by drying under a stream of nitrogen. Its enhancement factor (EF) was then determined to be 2.7x10$^7$ for the 1574 cm$^{-1}$ SERS mode assuming a value of 1.8x10$^{15}$ molecules/cm$^2$ for the surface concentration of chemisorbed benzenethiol.[20] It was recoated in April 2012 when the value of the SERS signal had decreased by an order-of-magnitude compared to its initial value. The substrate was rinsed in ethanol for 20 min prior to recoating using the same procedure as used initially.

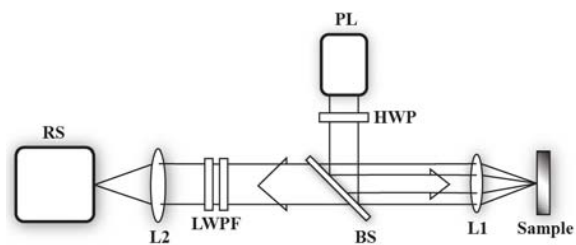

**Figure 1**. Schematic of the optical setup for the SERS measurements.

A schematic of the optical setup is shown in Fig. 1. The acronym PL is for pump laser. HWP is a half-wave plate for setting the polarization of the pump laser. BS is a beamsplitter reflecting the pump laser beam towards the SERS sample. L1 is a lens for focusing the pump laser to a 40±5-μm diameter spot on the sample and for the collimation of the SERS beam. LWPFs are long-wave-pass filters for blocking the pump light. L2 is a lens for focusing the SERS beam to the 80±10-μm spot on the entrance slit of the Raman spectrometer (RS). The pump laser is a 784.9-nm, 1400-mW cw single-mode (both transverse and longitudinal), linearly polarized Sacher Lasertechnik diode laser model SYS-420-0785-1400 (Sacher Lasertechnik, Marburg, Germany/Buena Park, CA). The lens L1 is a 12-mm-diameter, 20-mm-effective-focal-length (EFL), near-infrared (NIR) achromat. The long-wave-pass filters are 25-mm-diameter Semrock part number LP02-785RE-25 (Semrock, Rochester, NY). The lens L2 is a 25-mm-diameter, 40-mm-EFL, NIR achromat. The Raman spectrometer is a 500-mm-focal-length, f/6.5, Princeton Instruments model SP-2556 (Princeton Instruments, Trenton, NJ), which is equipped with the Princeton Instruments camera PIXIS 100BR. The grating in the spectrometer has 1200 grooves/mm. The blaze wavelength of the grating is 750 nm. The resolution of the Raman spectrometer for the 784.9-nm pump laser beam was determined to be 2.0±0.1 $cm^{-1}$.

RESULTS AND DISCUSSION

Three spectra for the 334-851, 818-1288, and 1281-1707 $cm^{-1}$ range were recorded after the recoat using 1.0-mW of pump power and 1.0-s integration time for each spectrum. On the second step the sample was heated for 3.5 min at 100º C and 1.5 min. at 125º C and three spectra for the same frequency ranges were recorded using the same laser setup. As a third step the sample was exposed for 10 s to the 40 mW of 785-nm pump laser power in 40±5 μm diameter spot, corresponding to pump laser intensity of 3.2 $kW/cm^2$ with the following measurement of the spectra. Figure 2 shows composite spectra for the total range 350-1700 $cm^{-1}$ for the all three steps.

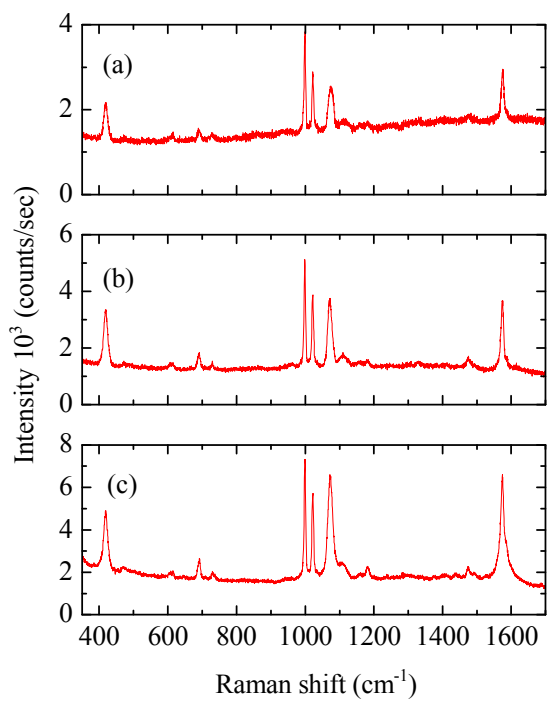

**Figure 2.** SERS spectra of benzenethiol on AgFON substrates measured from (a) recoated sample; (b) recoated sample after heating for 3.5 min at 100º C and 1.5 min. at 125º C; (c) recoated sample after heating and 10 sec exposure to 40 mW laser irradiation at 785 nm wavelength. The intensity scale is different for all three plots.

The main modifications in the spectra due to the heating and laser irradiation are: (a) the integrated peak intensities are enhanced after each step; (b) the relative intensities of the peaks change; and (c) the background intensity is modified. To analyze the changes in the peak intensities five fingerprint vibrational modes of neat benznethiol, $\omega_1$ at 415 cm$^{-1}$, $\omega_2$ at 1001 cm$^{-1}$, $\omega_3$ at 1026 cm$^{-1}$, $\omega_4$ at 1094 cm$^{-1}$, and $\omega_5$ at 1584 cm$^{-1}$ were used.

The average increase in the SERS signal upon heating is about two times for the five modes considered here. After the laser exposure the average SERS signal has increased additionally by 1.8 times. The values of the integrated SERS signal for post-heating and post-exposure relative to those for pre-heating for the modes $\omega_1 - \omega_5$ are listed in Table 1. The increase in the integrated SERS intensity upon heating and laser exposure is the largest for the $\omega_5$ mode and the smallest for the $\omega_2$ mode. The average value of the increase in the integrated SERS signal is 3.6. The SERS enhancement factors for the analyzed modes for post-heating and post-exposure were determined as listed in Table 1, assuming a value of 1.8x10$^{15}$ molecules/cm$^2$ for the surface density of chemisorbed benzenethiol molecules. The value of EF for the 1574 cm$^{-1}$ SERS mode, corresponding to $\omega_5$, had been determined to be 2.7x10$^7$ when the substrate was initially coated with benzenethiol. This implies that the surface density of benzenethiol molecules on the recoated sample after heating and laser exposure is $8 \times 10^{14}$ molecules/cm$^2$, which is 44% of the value of $1.8 \times 10^{15}$ molecules/cm$^3$ for the initially coated sample.

**Table 1**. The integrated SERS signals for post-heating and post-exposure relative to those for pre-heating and SERS enhancement factors (EF) for post-heating and post-exposure. The frequencies of the $\omega_1$, $\omega_2$, $\omega_3$, $\omega_4$, and $\omega_5$ are 415, 1001, 1026, 1094, and 1584 cm$^{-1}$, respectively.

| Mode | $\omega_1$ | $\omega_2$ | $\omega_3$ | $\omega_4$ | $\omega_5$ |
|---|---|---|---|---|---|
| SERS Signal | 3.4 | 2.4 | 3.3 | 4.1 | 4.9 |
| EF | $4.9 \times 10^6$ | $1.5 \times 10^6$ | $3.8 \times 10^6$ | $2.3 \times 10^7$ | $1.2 \times 10^7$ |

The relative intensities of the peaks for all three consequent steps are collected in Table 2. We notice that after the laser exposure the relative intensities of the Raman peaks are similar to the relative intensities measured for PhSH molecules chemisorbed on nanoengineered Au substrates.[15]

**Table 2**. The integrated intensities of Raman peaks relative to the $\omega_2$ mode at 1001 cm$^{-1}$ are given for three conditions of the recoated sample: pre-heated, post-heated (3.5 min at 100º C and 1.5 min. at 125º C), and post-heated and exposed (10 sec exposure to 40 mW of 785-nm laser power). The frequencies of the $\omega_1$, $\omega_3$, $\omega_4$, and $\omega_5$ modes are 415, 1026, 1094, and 1584 cm$^{-1}$, respectively.

| Mode | $\omega_1$ | $\omega_3$ | $\omega_4$ | $\omega_5$ |
|---|---|---|---|---|
| pre-heated | 0.75 | 0.56 | 1.36 | 1.03 |
| post-heated | 1.23 | 0.83 | 1.83 | 1.42 |
| exposed | 1.07 | 0.76 | 2.35 | 2.12 |

We hypothesize that during the recoating some of the benzenethiol molecules are adsorbed on the surface of the SERS substrate in intermediate or *precursor* states. These states may be conditioned by the presence of a partially ordered layer of chemisorbed benzenethiol molecules on the surface. Then the increase in the SERS intensity upon heating and laser exposure results from the conversion of the benzenethiol molecules in the precursor state to the chemisorbed state.

The relative enhancements and the frequency shifts of Raman modes upon adsorption on a plasmonic substrate can be associated with the modification of the electronic structure.[15] Due to the local nature of chemisorption effects,[21] this can be understood from the analysis of small metal-molecular complexes.[15] Among the analyzed Raman modes, $\omega_1$ - $\omega_5$, the mode $\omega_4$, which involves C–S bond stretch, experiences the largest frequency shift (about -20 cm$^{-1}$) and also the largest change in the relative intensity. We use this mode to characterize the interaction of PhSH with the metal. As compared to the lower frequency mode $\omega_1$, which is strongly affected by the near Ag environment, the $\omega_4$ mode changes may be better associated with the charge transfer excitations due to the S–Ag interaction.

Figure 3 shows the pre-heating, post-heating, and post-heating & post-exposure SERS spectra near the $\omega_4$ mode. The shape of the line in the pre-heated and post-heated samples, Figs. 3(a)-(b), are asymmetric while in the post-heating and post-exposure spectrum, Fig. 3(c), it is almost symmetric. In all spectra the line is sufficiently shifted from its position in neat PhSH, which supports the hypothesis that the measured signal is from the molecules adsorbed on the surface. The S–H bending mode of benzenethiol at 918 cm$^{-1}$ is not observed in all measured SERS spectra.[18] This further confirms that the molecules are chemisorbed. The lineshape variation reflects the distribution of the mode shifts. The shape of the lines in Figs. 3(a)-(b) can be fitted with two Gaussians centered at 1071 and 1080 cm$^{-1}$, while the peak in Fig. 3(c) can be fitted with a single Gaussian at 1073 cm$^{-1}$.

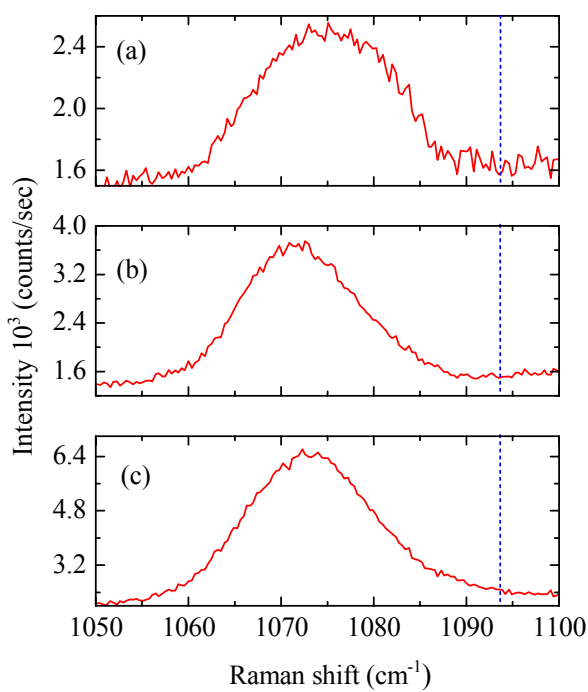

**Figure 3.** The lineshape of the $\omega_4$ Raman mode of benzenethiol at 1094 cm$^{-1}$ on AgFON SERS substrates measured from (a) recoated sample; (b) recoated sample after heating for 3.5 min at 100º C and 1.5 min. at 125º C; (c) recoated sample after heating and 10 sec exposure to 40 mW of 785-nm laser power. The dashed line shows the position of the peak measured in neat benzenethiol.

The shifts of the vibrational modes upon molecule chemisorption are associated with the modification of the ground electronic state, while the changes in Raman intensities also involve mixing of excited states localized in the metal and in the molecule. Thus, one should not expect a direct correlation between the mode shifts and modifications of Raman intensities. However, both effects stem from the electronic coupling between the molecule and the metal. The role of chemical interaction in SERS has be analyzed previously[22] using Albrecht's model of Raman scattering.[23] Far from resonance excitations the metal contribution in the Raman intensities is described by the B and C Albrecht's terms[22] that depend on the electronic transition dipole moments between the molecular and metal states. If there is no metal-molecule interaction this transition is forbidden, and the Raman intensities are not modified. On the other hand, if the states are mixed due to the electronic coupling the transition dipole would scale up with the mixture strength. This model provides a unified explanation of the observed distribution of mode shifts and the Raman intensities in the consequent experiments described above. After the recoating additional PhSH molecules are chemisorbed on the surface. However, their interaction with the surface is constrained by other molecules (including contaminants) that are adsorbed already. While the electromagnetic enhancement of the Raman signal for these molecules is the same as for the molecules forming a perfect monolayer, the chemical enhancement is weaker and also the shift of the $\omega_4$ mode at 1094 cm$^{-1}$ is smaller. After the structure is heated-up or irradiated by a strong laser field the molecules are rearranged[24] and a more homogeneous monolayer is formed.

It should be noticed that all Raman lines for the re-coated sample, Fig 2(a) can be assigned to PhSH molecules.[25,13,15] Neither of these lines disappear after the sample heating or laser exposure. This contrasts with Ref. 17 where several Raman lines were assigned to contaminants. Thus, we exclude the surface contamination by organic molecules as a reason for the spectral modifications. To confirm that the observed increase in the SERS signal with heating or exposure to a high-intensity laser field cannot be described by oxidation of PhSH on the silver surface we have computed Raman spectra of PhS–Ag$_9$ and PhSO–Ag$_9$ complexes shown in Fig. 4(a). Calculations were performed with Turbomole quantum chemistry package, version 5.10.[26,27] To compare the results with our previous calculations[13] we utilized the hybrid PBE0 functional.[28] The triple-$\zeta$ basis sets[29] and the split-valence basis sets with polarization[30] were used for the molecule and for Ag atoms respectively. The Raman spectra were computed for 2000 nm excitation wavelength to avoid contribution of metal-molecular resonances.[13] While the computed vibrational frequencies of benzenethiol are blue-shifted from the experimental values by about 20-40 cm$^{-1}$, the relative shifts of the modes due to adsorption are consistent with the measured values. The oxidation of the complex results in the quenching of some and in the formation of an additional vibrational mode in the range between the PhSH modes $\omega_3$ and $\omega_4$, which was not observed in the experimental spectra. Similar effects were obtained for other computed complex geometries. In addition, we have confirmed that the modes associated with the sulfur oxidation analyzed in Ref. 31 were not observed in our measurements either.

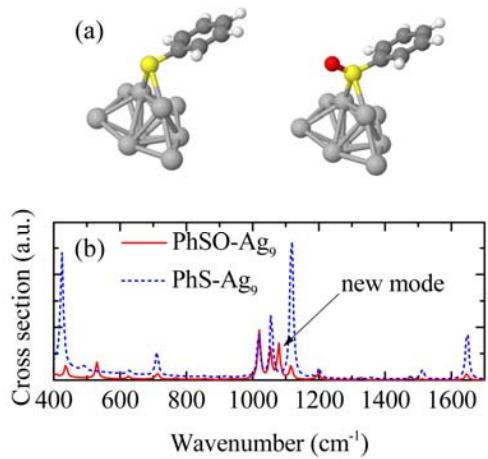

**Figure 4**. (a) Metal-molecule complexes PhS–Ag$_9$ and PhSO–Ag$_9$; (b) Raman spectra computed at 2000 nm excitation wavelength. Additional vibrational modes are associated with the oxidation of the complex.

CONCLUSIONS

We have observed about 100% increase in the SERS signal from benzenethiol recoated on a AgFON substrate upon heating at 100-125º C or exposure to high-intensity (3.2 kW/cm$^2$) 785-nm pump laser. The increase in the integrated SERS signal is accompanied by the modifications of relative intensities and shapes of fingerprint Raman modes. The observed effect may be understood as a conversion of the benzenethiol molecules, initially adsorbed in a precursor state, to a chemisorbed state caused by thermal and photo-thermal effects.

**Corresponding Author:** *R.L. Aggarwal, MIT Lincoln Laboratory, Lexington, MA 02420-9108, USA.  E-mail: *aggarwal@ll.mit.edu*

**Funding Sources**

The MIT Lincoln Laboratory portion of this work was sponsored by the Defense Advanced Research Projects Agency under Air Force Contract FA8721-05-C-0002.  The work at Harvard University was sponsored by the Defense Threat Reduction Agency under Contract No. HDTRA1-10-1-0046.

ACKNOWLEDGMENT

We thank Dr. T.Y. Fan, Dr. D. Rappoport, Dr. R. Olivares-Amaya and Prof. A. Aspuru-Guzik for their comments regarding this work, N.G. Greeneltch and Prof. R.P. Van Duyne for providing the SERS substrate, and Russell Goodman for depositing the benzenethiol SAM on the SERS substrate. The MIT Lincoln Laboratory portion of this work was sponsored by the Defense Advanced Research Projects Agency under Air Force Contract FA8721-05-C-0002.  The work at Harvard University was sponsored by the Defense Threat Reduction Agency under Contract No. HDTRA1-10-1-0046. Opinions, interpretation, and recommendations are those of the authors, and do not necessarily represent the view of the United States Government.